\documentclass[aps,a4paper]{revtex4}

\usepackage[utf8]{inputenc}
\usepackage{epsfig}
\usepackage{graphicx}
\usepackage{amsmath,amssymb,color}
\usepackage{float}
\usepackage[english]{babel}

\parskip=\medskipamount

\newcommand{\fig}[1]{Fig. \ref{#1}}

\newcommand{\be}{\begin{equation}}
\newcommand{\ee}{\end{equation}}

\newcommand{\barr}{\begin{array}}
\newcommand{\earr}{\end{array}}

\newcommand{\beqn}{\begin{eqnarray}}
\newcommand{\eeqn}{\end{eqnarray}}

\newcommand{\bs}{\begin{subequations}}
\newcommand{\es}{\end{subequations}}

\newcommand{\bw}{\begin{widetext}}
\newcommand{\ew}{\end{widetext}}

\begin{document}

\title{Did smartphones break the world as we knew it?}
\author{Mikhail V. Tamm$^{1}$}

\affiliation{$^1$ School of Digital Technologies, Tallinn University, Tallinn, Estonia}

\date{\today}

\begin{abstract}
I overview data on several radical societal changes started circa 2015: accelerated decline in fertility rate, backsliding of democracy, rise of populist politics and arrest in generational renewal of political leadership. I conjecture that all these processes have a common underlying cause: the spread of cheap and easy access to information due to wide spread of smartphones. I speculate about possible mechanisms connecting the observed changes with this underlying information revolution and discuss relevant historical parallels.
\end{abstract}

\maketitle

For decades we have been hearing politicians and current events commentators claiming that we are living in a world of innovation, where change is coming ever faster. In particular, the large emphasis has been put on the impact of the information revolution, which was supposed to produce incredible gains in productivity and dramatically change the structure of the societies. Meanwhile, this claims where in stark contrast with the actual evidence: from mid-1970s to mid-2010s the economic growth in the world-leading countries was much slower than in the previous decades and was slowing down throughout the period, and, except for the collapse of the communist system in Eastern Europe in 1989-91, the political and social structures were remarkably stable. It is understandable therefore that to many, including the author, this talk started to sound a lot like a boy crying ``Wolf!"  

\begin{figure}
    \centering
    \includegraphics[width=10cm]{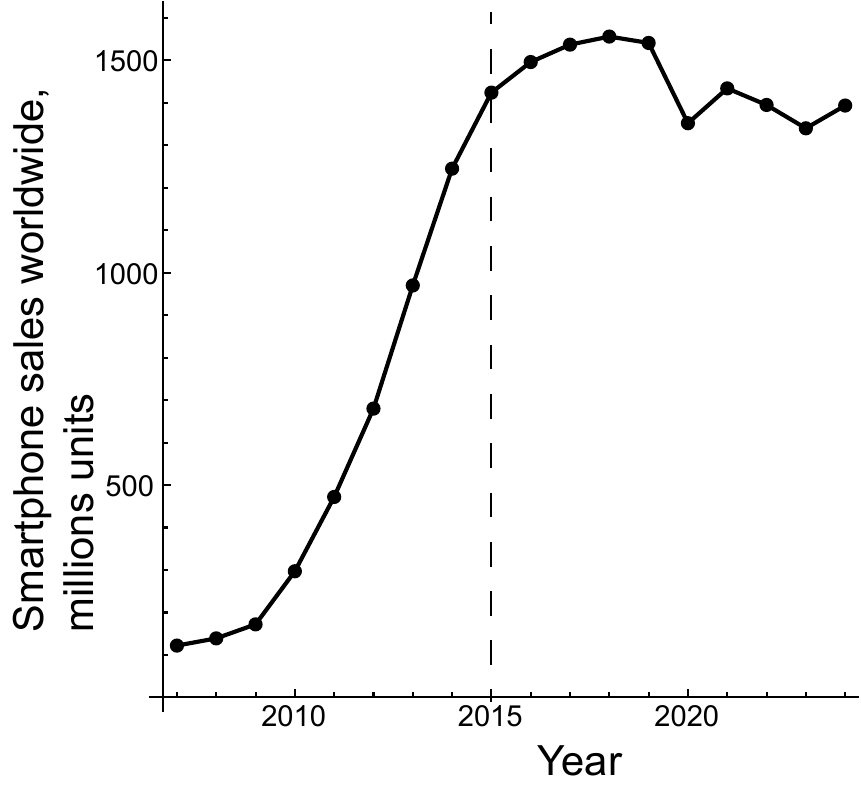}
    \caption{ 
Sells of smartphones worldwide \cite{statista} (the 2024 point is an approximate preliminary estimate according to \cite{counterpoint}). Note the change of dynamics from explosive growth to saturation around 2015 or slightly before.
    }
    \label{fig:1}
\end{figure}

However, there seems to be strong evidence that the long-predicted information revolution has finally came, and the form it took is not exactly like what we expected. This revolution has a rather clear starting point around 2015 (give or take a year), which is the point in time when smartphones transitioned from a toy for early adopters to a universally accessible product (see \fig{fig:1}). Numerous indicators, which were relatively stable before 2015, have since begun shifting visibly and continuously. Importantly, this shifts continue to the present day without any evidence of saturation. Putting it in other words, the smartphone revolution is still ongoing: the societies around the world have clearly departed from their pre-2015 norms, but they are yet to fully adsorb the ``smartphone shock''.

In what follows we inspect in some depth two particular groups of such shifting indicators, one related to the demography, another --- to the state of political systems around the globe.

\section{Birth rates}

\begin{figure}
    \centering
    \includegraphics[width=17cm]{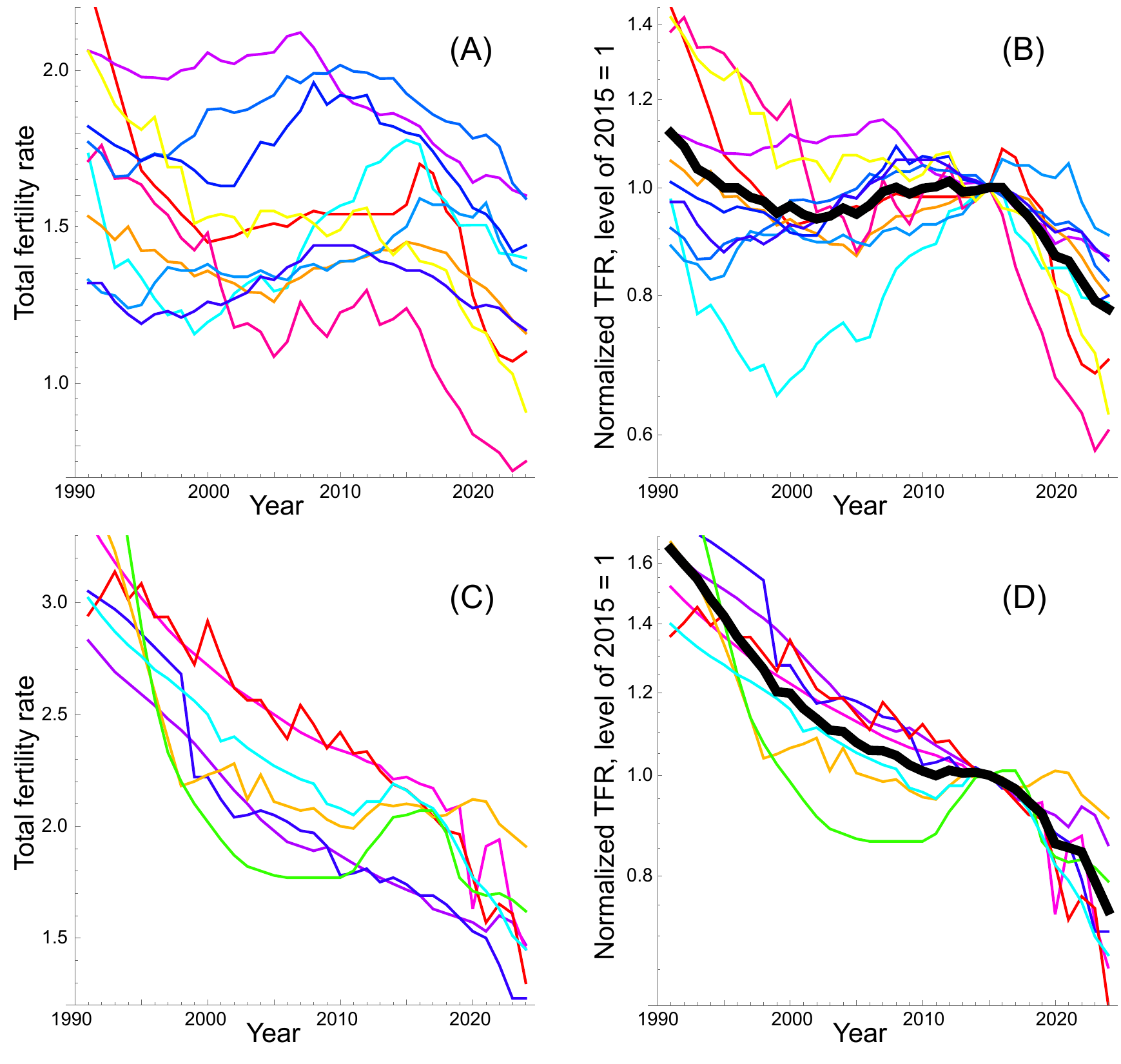}
    \caption{ 
Total fertility rates (TFR) in large (populations more than 50M) countries with low fertility. (A),(B): countries where transition to low fertility mostly finished by early 1990s: China (red), Japan (orange), South Korea (magenta), Thailand (yellow), USA (purple), Russia (cyan) and 4 big Western European countries (bluish colors, from lighter to darker: Germany, France, UK, Italy); (C), (D): countries where transition to low fertility happened by 2020: Brazil (purple), Mexico (magenta), Colombia (blue), Philippines (red), Vietnam (orange), Iran (green) and Turkey (cyan). Panels (A) and (C) show the raw values of TFR, panels (B) and (D) show TFRs normalized to their level in 2015; thick black lines in panels (B) and (D) are averages over the corresponding country groups. Note the rapid decline in fertility rates in both groups after 2015.
    }
    \label{fig:2}
\end{figure}

One aspect to point out is a significantly accelerated decline in birth rates in recent years. \fig{fig:2} shows the data for the total fertility rates (TFR, the average number of children a woman is expected to bear in a lifetime) in the largest countries of the world in the last 35 years. 

To understand what is going on, we should first recall the notion of the demographic transition\cite{landry,davis,notestein}, i.e., the shift from high to low birth rates which is driven, most importantly, by radical decline in child mortality, growth of women education and onset of social safety nets. This transition is universal but it occurred in different countries at different points in time, e.g., in France it was well underway by 1900, while in Ethiopia it is still only starting. Thus, for a meaningful cross-country comparison of the dynamics of birth rates it is essential to group countries according to the stage of the demographic transition they are on. 

We restrict ourselves to the countries in the advanced stages of the demographic transition. Indeed, the countries on the early stages typically show a very rapid transition-related decrease of the TFR, and it is hard to distinguish any secondary effects on top of this secular trend, especially given that demographic data for many such countries is of rather low quality.  Out of 30 countries with population above 50 millions residents, 10 (China, USA, Russia, Japan, Germany, France, UK, Thailand, Italy and South Korea) had their demographic transition mostly finished by early 1990s, while 7 others (Brazil, Mexico, Philippines, Vietnam, Turkey, Iran and Colombia) finished their transition between 1990 and 2020. In  \fig{fig:2} we present the TFR data for these two groups of countries separately. Most importantly, thick black lines in \fig{fig:2} (B) and (D) are the group averages of the TFR. 

It is notable, that TFR for the countries after the demographic transition was mostly stable in the 20 years before 2016 (and some decline in the early 1990s is mostly due to the final stages of demographic transition in China, Thailand and South Korea), while after 2016 the plateau changed to a rapid and approximately linear decline, with average TFR falling by more than 20\% in 9 years, from approximately 1.75 to approximately 1.35. 

Similar trend is also evident in the the behavior of the TFR in the countries in the late stage of demographic transition (see \fig{fig:2} (D)): the ``normal'' decline in the TFR due to the demographic transition, which was approaching saturation in early 2010s, is replaced after 2016 with a much faster decay.

Notably, while the phenomenon of demographic transition is widely known and well understood and their is plenty of literature on the subject, this recent additional drop in fertility is much less discussed and understood. It is sometimes believed that over-worrying about low fertility rates is just a pet topic for hard right politicians, and that in fact the societies can well adapt to low fertility. It might be true for the rates typical for most rich countries 10 years ago, but much less so if the fertility rates remain at their current level or drop even further. Indeed, a simple back-of-the-envelope calculation shows that, assuming absence of migration 30 years gap between generations and 0.25\% per year growth in the life expectancy, the stationary (in terms of the shape of the population pyramid) state with TFR = 1.75 corresponds to population decrease of 0.28\% per year or 25\% per century, while TFR = 1.35 corresponds to population decrease of 1.15\% per year, which in a century results in 3-fold decrease in population. The first of these two regimes is hardly problematic: the population decrease on this scale is manageable, and can actually be reversed by a moderate level of immigration, corresponding to stationary level of foreign-born population at 10-15\% or so, which is similar to historical levels in USA and some European countries. Meanwhile, the second regime is quite scary indeed: in the absence of migration it leads to catastrophic levels of elderly dependency ratio, and the level of migration needed to stabilize the population implies a foreign-born population close to 50\%, making assimilation of migrants essentially impossible \cite{footnote1}.

It is not completely clear what is the reason behind the recent downward trend in fertility. However, the fact that it is so universal (e.g., in all 10 countries in  \fig{fig:2} (A,B) the fertility dropped between 2015 and 2024, while it grown in 8 of them between 2005 and 2015 and in 5 of them between 1995 and 2005) and so synchronous suggests that the reason behind it should also be universal and have a clear onset. The world-wide penetration of smartphones seems a reasonable candidate. Indeed, there is abundant literature on the behavioral changes related to the spread of online social networking. These changes include ever-growing fraction of romantic partners meeting each other online\cite{rth}, growth of the fraction of people who are not and never have been in a romantic relationship\cite{ft}, decreased sexual activity \cite{pvi, stone}, etc. Thus, internetization has radically changed and continues to change the whole set of norms and rituals around sexual behavior and family formation. It seems only natural to assume that it might influence fertility as well.

Importantly, the observed drop in fertility is seemingly not yet reached saturation, which might be explained by the slow renewal of the child-bearing generations with people who came of age in the smartphone era gradually replacing the older cohorts.


\section{Political system}

\begin{figure}
    \centering
    \includegraphics[width=17cm]{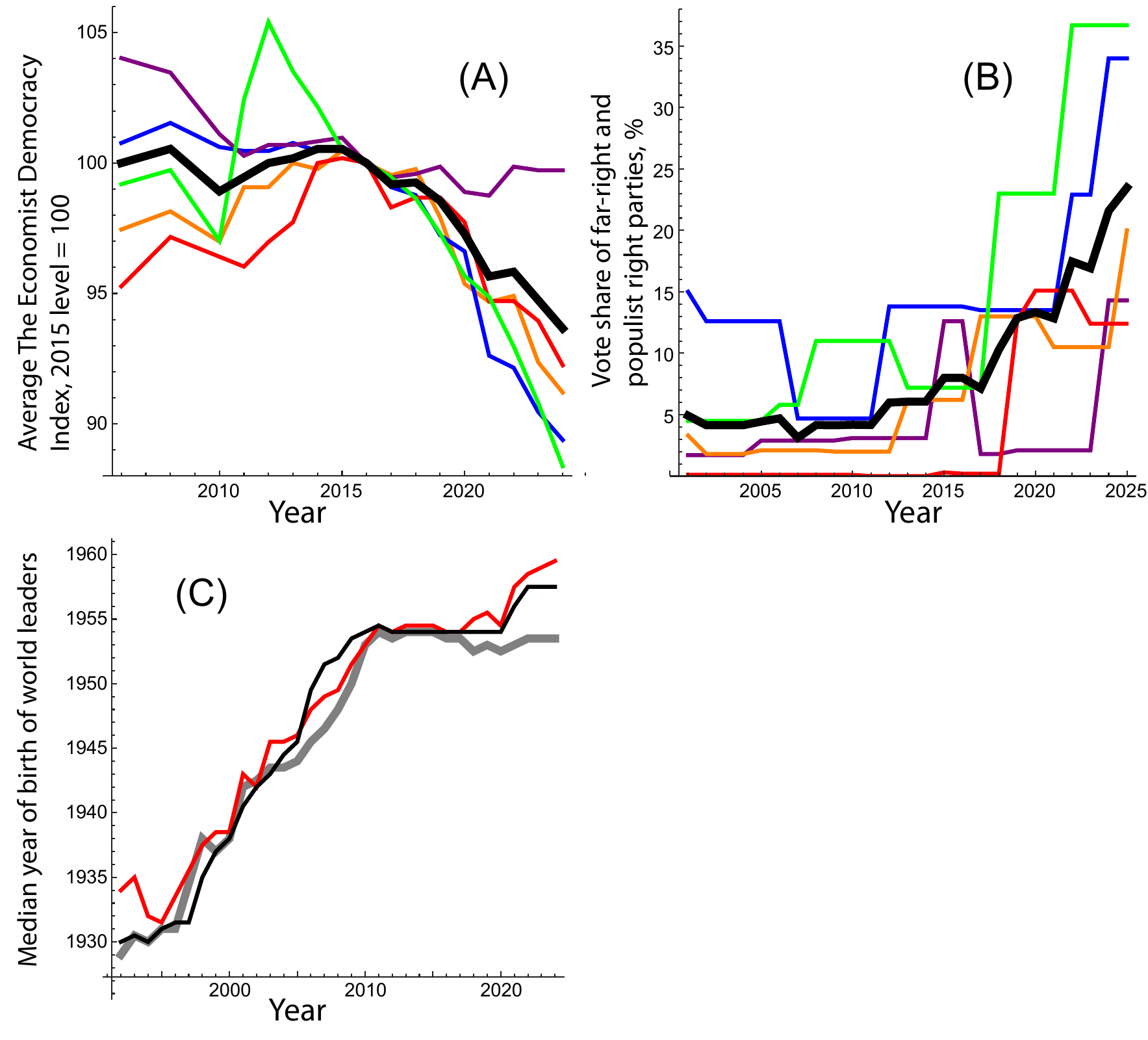}
    \caption{ 
    Three plots illustrating the change in the state of political system world-wide. 
    Panel A: Democratic backslide. Average value of The Economist Democracy Index worldwide (thick black line) and in America (blue), Asia (red), Europe (purple), Middle East and North Africa (green) and sub-Saharan Africa (orange). Each curve is normalized to its level in 2015. Note that relative stability before 2015 is a result of averaging over different trends in different regions, while after 2015 there is a world-wide downward trend.
    Panel B: Populist revolution. Vote share of right-populist and far-right parties in the most recent legislative election in Germany (orange), France (blue), the UK (purple), Italy (green) and Spain (red)\cite{footnote}; thick black line is the average over these 5 countries.
    Panel C: Clotted leadership renewal. Median year of birth leaders of 20 most populous countries of the world (thick gray line), of G20 countries (black line), of 24 big (population>20 mln) democracies. Democracies here are countries which at any year between 2015 and 2024 had The Economist Democracy Index 6.00 or larger, namely Argentina, Brazil, Canada, Colombia, Mexico, Peru, USA, Australia, India, Indonesia, Japan, Malaysia, Philippines, Taiwan, Thailand, South Korea, France, Germany, Italy, Poland, Spain, UK, Ghana and South Africa.}
    \label{fig:3}
\end{figure}

If we accept that the smartphone revolution can measurably influence such fundamental things as the number of children women decide to have, it is no surprise that it might fundamentally upend more fluid properties of the society like political systems.

One often discussed aspect is the backsliding of democracy observed worldwide in the last 10 years (see \fig{fig:3} (A)). Note that while prior to 2015 different regions where drifting in different directions, the downward trend in the last 10 years is universal (albeit somewhat less pronounced in Europe). 

Another aspect is the increase in popularity of populist right and far right parties, which is illustrated in \fig{fig:3} (B) by the data for 5 big Western European democracies: Germany, France, UK, Italy and Spain. Up to mid-2010s the vote share of right populist and far right parties was relatively stable ar around 4-5\% while since then it seems to be growing steadily by 1.5\% per year or even more without any sign of saturation. The data presented for each country is for the last parliamentary election, thus the mean is effectively averaging over the last parliamentary term of $\sim 4$ years. The exact position of inflection point is up to debate, but by 2017 the upward trend is clearly seen, implying an inflection somewhere in the previous (2013-17) parliamentary term.

Finally, rather perplexingly, many countries the last 10-15 years have seen a breakdown in generational renewal of political leaders. In \fig{fig:3} (C) we show the median year of birth of the leaders of 20 largest by population countries in the world (gray), countries-members of G20 club (black) and 24 largest democracies (defined as countries with the Economist Democracy Index above 6.0 at any point between 2015 and 2024, red). Clearly, in all 3 groups of countries there is a marked arrest of generational renewal in the 2010s, which is partly continuing to this day. In the group of 20 largest countries the median age of leaders has grown by 13.5 years between 2011 and 2024, in 24 democracies it grown by 9 years between 2011 and 2020 (since then it declined by 1 year but is still 5 years above the 1992-2011 average).

The rise of populist parties and democratic backslide worldwide are often attributed to either economic problems (slow growth, growing inequality, loss of prestige of blue-collar jobs) or to the rapid growth of migration from poorer to richer countries. Notwithstanding the fact that both these problems are clearly important in many countries, they seem unable to explain the world-wide nature of the phenomena. Indeed, for example, India and Poland have seen significant growth of populist politics simultaneously with stellar economic performance, while Donald Trump was first elected President of the United States on the anti-immigration platform despite lowest level of illegal immigration in a generation. Reaction to a supposed excess of far-left populism is another often-cited reason. In fact, although left-wing populism has grown in popularity simultaneously with the right-wing sort, and it does have certain influence here and there, its electoral success is much less prominent (with the arguable exception of Mexico).

We suggest that worldwide and almost synchronous nature of all three aforementioned trends is better understood if we think of them as side-effects of the information ("smartphone") revolution, which happened almost simultaneously around the world. In what follows we suggest possible causal mechanisms relating radical democratization of information access to the observed phenomena and discuss possible historical analogues of the revolution we are living through.

The idea that democracy is by nature susceptible to a takeover by a populist strongman is well understood since antiquity, and at least since the Republican Rome the (not always successful) solution to this problem was to create institutions and traditions which channel and regularize the will of the people and prevent a leader, regardless of the level of his support, to do whatever he wants. Wherever these checks do not exist and/or are destroyed, the collapse of democracy is usually imminent.

The details of such checks, rooted partly in law and partly in tradition, are specific to each democracy and might change with changing  historical circumstances, including the prevalent levels of technology. It is clear that in the post-war era both traditional media and political parties played essential role in these checks. The norms of the public discourse defined by the media included fact-checking and calling out politicians lying publicly and promoting overheated emotional rhetoric. Political parties worked as gate-keepers \cite{howdemdie} preventing the takeover of political space by demagogue politicians (George Wallace or Pat Buchanan in the US can be good examples).  

The radical democratization of the information space means that norms developed by traditional media are very easy to circumvent: the new, wider info-sphere of blogs, podcasts, influencers, messenger groups, etc., on the one hand, provides audiences with a much larger volume of much more diverse information, and, on the other hand, has no comparable set of norms and traditions. At least short- and medium-term it leads to multiple adverse outcomes.  

First, on the individual level people have to prioritize selecting and filtering information over analyzing it, and humans are generally not that good in this task\cite{calling_bullshit}. As a result, it has been shown again and again \cite{delvicario,cinelli} that audiences self-sort to consume news and information, which is confirming their pre-existing biases, self-radicalize by consuming emotionally charged material and often show no interest in distinguishing between real and fabricated information, provided it confirms their biases and grievances.

Second, all sorts of existing and would-be authoritarian leaders use new methods of surveillance and control to suppress dissent, interfere in the information landscape, corrupt politicians and institutions of the democratic world and use increased connectivity to exchange ``best practices'' between each other \cite{autinc}.

Third, democratization of information landscape leads to dramatic lowering of entrance barriers to politics. This leads to political parties losing their gate-keeping function, i.e. losing the ability to select the new governing elite. In many cases (Democratic party primary of 2020 with its 27 'major' candidates might be a good example) the choice becomes so wide that a little-known but traditionally competent candidate has no hope of getting noticed over multitude of extravagant and downright extreme alternatives. Thus, leadership is systematically vested either upon populist politicians (be it by hostile takeover of traditional parties or by the growth of new ones) or upon well-known elder statesmen of the pre-2015 era presenting themselves as the only shield against the populist uprise. In fact, on the populist flank the dynamics is similar: in order to get noticed in the over-competitive field, a populist leader has to have some original core of support, which leads to the advancement of elderly fringe politicians who spent decades on the periphery of the political landscape. Taken together, this explains the apparent stoppage of the leadership generational renewal. 

This stoppage is, obviously, a transient phenomenon: inevitable aging of the leaders will eventually ensure renewal, and there is evidence (see \fig{fig:3}C) that this is already happening. Meanwhile, the decay of the influence of traditional political parties is a more gradual and long-term trend: triggered by information revolution 10 years ago it is evolving slowly but inexorably, and arguably is the explanation behind similarly inexorable growth of right wing populists' voting share in Europe and beyond. 

\section{Discussion}

We thus argue that the reasons behind the current crisis of democracy is less a result of particular economic and social imbalances (those imbalances, of course, do exist and should be addressed, but this is as true now as it was 20 or 50 years ago) but rather a structural effect of the development of the information technologies. Why is this conclusion important?

For one thing, it reduces the desire to play a blame game. Yes, political and cultural elites are sometimes self-serving and condescending. But historically, this is true at all times, and the new populist elites, which electorates seem to be eager to put in place of the old ones, are even more shamelessly self-serving and condescending.  
Yes, current leadership of traditional liberal-democratic parties is depressingly uninspiring. But this is not the reason of the current crisis, it is its symptom: advancement of high-quality leaders in traditional parties is failing exactly because of the structural changes brought in by the information revolution. And the politicians on the left and right populist extremes are not ``natural counter-reactions'' to one another but different sides of the same populist medal, whose raise to prominence is governed by exactly same underlying forces.

Second, it suggests some instructive historical parallels. Current is not the first information revolution in the history of the mankind, and while we know relatively little, e.g., about historical consequences of the invention of writing, the consequences of the two more recent information revolutions -- the inventions of printing press and radio are much more well-documented.

Similarly to the current one, these previous information revolutions where all about democratization of access. According to some counts, in the 1520s there was roughly million of copies of Martin Luther's pamphlets in circulation\cite{edwards}, without a printing press a number three orders of magnitude smaller would hardly be achievable. Radio (and to less extent newsreels, a more or less simultaneous bi-product of the invention of cinema) for the first time made current events and political discussions on the national scale accessible to all. 

There is no doubt that inventions of printing press, radio and cinema where in the long run extremely beneficial to humanity. Development of science as a social institution would not be possible without printing. Rapid exchange of technical knowledge in printed form was essential for the industrial revolution. Affluent society with modern safety nets is a result of wide democratic participation made possible by radio, cinema and TV. 

It is hard not to notice, however, that the onset of these inventions was accompanied by political crises of extraordinary scale: a century of religious wars all over Western Europe in the first case, and the horrors of totalitarian regimes and World War II in the second\cite{footnote3}. Clearly, this is not a coincidence but rather a result of political discourse being successfully hijacked by the rabble-rousing demagogues in a striking similarity to the current political dynamics. 

Eventually, by the second half of the 17th century the Western European societies did find new ways to muddle through by accepting some basic guarantees of personal liberties (most importantly, religious tolerance) and, in case of Britain after the Glorious revolution and later the USA, the parliamentary regime with separation of powers. Similarly, after the World War II the modern liberal democratic world order was established, implying multilateral institutions and at least superficial adherence to non-aggression and human rights on the international level, and norm-based policing of the discourse by traditional media and political parties on the intra-state level. These solutions where, of course, always messy and imperfect, and it it notable that they were only accepted as a lesser evil by societies in fear of the return to the large-scale violence of the transition years. 

There is no doubt that ways of suppressing demagoguery in the new information era can also be found. A total lack of advancement in this field for the last 10 years seems to come not from the lack of ideas but from the lack of will: general public is seemingly completely disinterested in the subject. Historical precedent suggest that the only way for it to change its mind is to become properly scared by the alternative. If this is true, we are up to a very bleak time indeed.


\section*{Acknowledgements}
I am grateful to A. Raksha, who coined the idea of section I to me in a private conversation.

\end{document}